\documentclass[aps,pra,twocolumn,amsmath,amssymb,nofootinbib,showpacs,superscriptaddress]{revtex4-1}
\usepackage[english]{babel}
\usepackage{float}
\usepackage{latexsym}
\usepackage{graphics}
\usepackage{graphicx}
\usepackage{epsfig}
\usepackage{color}
\usepackage{bm}
\usepackage{amsmath}
\usepackage{amssymb}
\usepackage{amsthm}
\usepackage{dcolumn}
\usepackage{bm}
\usepackage{float}
\usepackage{hyperref}
\usepackage{color}
\usepackage{epstopdf}
\usepackage{braket}
\usepackage{cleveref}
\usepackage{tabularx}
\usepackage{multirow}
\usepackage{booktabs}
\usepackage[svgnames]{xcolor}
\hypersetup{hidelinks,colorlinks=true,allcolors=DarkBlue}

\DeclareMathOperator*{\argmin}{arg\,min} 

\theoremstyle{remark}

\begin{document}

\preprint{APS/123-QED}

\title{Simulating methylamine using symmetry adapted \\ qubit-excitation-based variational quantum eigensolver}

\author{Konstantin M. Makushin}
\affiliation{Kazan Federal University, Kazan 420008, Russia}
 \email{llconstantinel@gmail.com}

\author{Aleksey K. Fedorov}
\affiliation{National University of Science and Technology ``MISIS'', 119049 Moscow, Russia}

\date{\today}
\begin{abstract}
Understanding the capabilities of quantum computer devices and estimating the required resources to solve realistic tasks are among the most important problems towards achieving useful quantum computational advantage.
As the conventional realization of the variational quantum eigensolver (VQE) approach requires a significant amount of resources, we propose and analyze optimization strategies that combine various methods, including molecular point group symmetries (symmetry adaptation), compact excitation circuits (qubit-excitation-based), different types of excitation sets, and qubit tapering.
These strategies allow for a significant reduction in computational requirements while ensuring convergence to the correct energies.
First, we apply these combinations
to small molecules, such as LiH and BeH$_2$, to evaluate their compatibility, accuracy, and potential applicability to larger problems.
We then simulate the methylamine
molecule within its restricted active space using the best-performing optimization strategies. 
Finally, we complete our analysis by estimating the resources required for full
active-space simulations of the methylamine and formic acid molecules. 
Our best-performing optimization strategy reduces the number of two-qubit operations for simulating the methylamine molecule from 600,000 (in the STO-3G basis with a naive Unitary Coupled Cluster ansatz) to approximately 12,000, using 26 qubits.
Thus, the proposed combination of optimization methods can reduce the number of two-qubit operations by nearly two orders of magnitude. 
Although, we present alternative approaches that are of interest in the context of the further optimization in the number of two-qubit operation, we note that these approaches do not perform well enough in terms of the convergence to required energies.   
While these challenges persist, our resource analysis represents a valuable step towards the practical use of quantum computers and the development of better methods for optimizing computing resources.
\end{abstract}

\maketitle

\section{Introduction}

The problem of calculating energies and electronic structures of molecular ground and excited states is of primary importance for various industry-relevant problems, 
such as finding catalysts for chemical reactions and materials design~\cite{Helgaker2000}.
However, the computational complexity of exact quantum chemistry simulations for many relevant situations grows exponentially with the system size~\cite{Kitaev2006}.
This problem has stimulated the development of classical approaches that capture relevant chemical properties~\cite{Gunsteren2024}, 
yet used simplifications may hinder some underlying electronic correlation effects that may play important role for various cases. 

To overcome these limitations and address the problem of growing computational demands, the quantum computing paradigm presents a promising solution~\cite{Lloyd1996,Cao2019,Aspuru-Guzik2020}. 
By leveraging the principles of quantum physics, quantum computing devices offer the potential to efficiently solve complex quantum-chemical problems that are intractable for classical computers~\cite{Cao2019,Aspuru-Guzik2020}.

There are several approaches to utilize the resources of quantum computing devices to solve computational problems in quantum chemistry. 
First, special-purpose quantum devices can be used~\cite{Fedorov2022}. 
In particular, boson sampling~\cite{Aspuru-Guzik2015}, analog quantum simulators~\cite{Cirac2019}, and quantum annealing~\cite{Leib2019} devices have been used for solving various chemical problems; 
however, encoding of chemical problems to the setting of special-purpose quantum computers mays hinder the potential quantum computational advantage~\cite{Xia2018,Chermoshentsev2022}. 

Second, universal quantum computing models, specifically, gate-based quantum computers, have the potential to be used for quantum chemistry simulation~\cite{Lloyd1996,Cao2019,Aspuru-Guzik2020}. 
In order to encode electronic states of molecules into qubits, one uses a basis of predefined \textit{spin-orbitals}, i.e., quantum states that can be occupied by individual electrons.
Each qubit in the quantum register represents one such spin-orbital, with its value (0 or 1) determining whether this spin-orbital is occupied or not. 
The molecular state is then a superposition of multiple occupancy configurations. 
The quantum computing device is then used, e.g., to find the lowest-energy state of molecules. 

However, the full quantum simulation using gate-based quantum computers require prohibitively large amount of resources, specifically, the number of two-qubit operations~\cite{Cao2019,Aspuru-Guzik2020}. 
The ability to implement two-qubit operations in existing quantum computing devices is limited by the amount of errors due to the effects related to interactions with environment. 
In this context, the variational quantum eigensolver (VQE) approach~\cite{Aspuru-Guzik2014,Aspuru-Guzik2016,Babbush2021-4,Sapova2022} 
is of interest since this algorithm can be implemented already with the use of existing noisy intermediate-scale quantum (NISQ) devices. 
VQE has shown its ability to calculate molecular ground states and energy spectra with chemical accuracy in a number of studies~\cite{Babbush2021-4,Sapova2022,Fedorov2022,Tilly2022}. 
An overview of the developments of VQE showcasing its adaptability and potential for complex quantum systems is presented in Ref.~\cite{Fedorov2022}.
For example, the recent study~\cite{Tilly2022} has highlighted the methodological advancements and best practices in applying VQE to quantum chemistry problems, which emphasises its versatility for a variety of molecular systems.
VQE has been used in benchmarking studies on a variety of quantum hardware platforms and molecular configurations, see Refs.~\cite{bentellis2023benchmarking,hu2022benchmarking}.
Applications of VQE in solving specific tasks, such as applying sophisticated Hamiltonian corrections for molecular simulations or calculating the magnetic properties of rare earth ions, has been demonstrated in Refs.~\cite{Singh2024,Makushin2023}. 
Large-scale implementations on high-performance supercomputers, which highlights the scalability of VQE for difficult systems, has been studied in Ref.~\cite{Shang2022}.
Recently proposed improvements in the VQE approach include improvements in circuit depth optimization and error mitigation~\cite{Steiger2024} and
further reduce computational expenses by using symmetry-compressed double factorization approaches~\cite{Rocca2024}.
All of this research supports the idea that VQE is an option for solving quantum-chemical issues that are beyond the capabilities of classical computers, when combined with advancements in algorithmic technology and better quantum hardware.
Variational approaches based on VQE have been used to analyze small molecules, 
such as hydrogen (H$_2$)~\cite{Gambetta2017-2,Martinis2016-2,Blatt2018-3}, lithium hydride (LiH)~\cite{Gambetta2017-2,Blatt2018-3}, beryllium hydride (BeH$_2$)~\cite{Gambetta2017-2}, 
and water (H$_2$O)~\cite{Monroe2020-2}, 
as well as to simulate diazene isomerizations~\cite{Babbush2020} and carbon monoxide oxidation~\cite{Sapova2022}. 

Methylamine (CH$_{3}$NH$_{2}$) and formic acid (CH$_2$O$_2$), also known as methanoic acid, are both simple yet fundamental organic compounds in chemistry and biochemistry. 
Methylamine, an amine related to ammonia, is crucial in manufacturing pesticides, pharmaceuticals, and solvents and constitutes amino acids, which are building blocks of protein~\cite{Wu2021}. 
Formic acid, the simplest carboxylic acid, is found in insect stings and is utilized industrially in making leather and textiles, and as a feed preservative. 
Both play significant roles in biological systems~\cite{fedorov2021towards}, with formic acid also acting as a key metabolic agent and a strong solvent.
More precise understanding of their properties and features of compounds on their basis may provide valuable information for various problems chemistry and life sciences. 

In this work, we analyze the resources that are required to simulate certain molecules on a medium-scale quantum computer with the use of the VQE approach, taking into account possible optimization.
Specifically, we combine the use of molecular point group symmetries (symmetry adaptation), compact excitation circuits (qubit-excitation-based), several types of excitation sets, and the qubit tapering method.
As we demonstrate, these optimizations allow for a significant reduction in computational requirements while ensuring the convergence to the required energies.
First, we apply these combinations to small molecules, such as LiH and BeH$_2$, to evaluate their compatibility, accuracy, and potential applicability to larger problems. 
We then simulate the methylamine molecule within its restricted active space using the best-performing optimization strategies.
Finally, we complete our analysis by estimating the resources required for full active-space simulations of the methylamine and formic acid molecules, which, using our best optimization strategy, would require around 10,000–15,000 two-qubit gates. Both the two-qubit gate count and circuit depth are improved by nearly two orders of magnitude compared to the naive approach.
Our results demonstrate that these optimization strategies perform well and can make such simulations more feasible for near-term quantum computers.

The paper is organized as follows.
In Sec.~\ref{sec:VQE}, we formulate the VQE approach for solving quantum-chemical problems. 
In Sec.~\ref{sec:Resource}, we analyze the required quantum resourced for simulating methylamine and formic acid molecules using the VQE approach.
We take into account possible optimization techniques that can be used to reduce the amount of the required results for VQE-based quantum chemical simulations.
We summarize our results in Sec.\ref{sec:results} and present our conclusions in Sec.\ref{sec:conclusion}.

\section{VQE formalism}\label{sec:VQE}

During the last decade, a large variety of modifications of the VQE approach have been proposed~\cite{Babbush2021-4}, the majority of them is based on the Ritz variational principle.
The Ritz principle involves choosing a certain parametrized combination of wave functions or state vectors i.e. the ansatz, and minimizing the functional defined on this parametrized combination. 
Choosing the right ansatz should take into account the specifics of the problem and its boundary conditions. 
In other words, the ansatz represents a linear combination of some known functions, which are parameterized by some unknown coefficients:
\begin{equation}\label{ansatz_1}
	\Psi = \sum_{i}\theta_{i}\psi_{i}.
\end{equation}

Let $\mathcal{H}$ be the Hamiltonian of a certain physical system, and its spectrum of eigenvalues is bounded from below by the minimum eigenvalue $E_0$. 
Assuming that the state vector $|\Psi\rangle$ can be described using a set of variable parameters $\boldsymbol{\theta}=\{\theta_{1}, \theta_{2}, \dots, \theta_{i}\}$ similarly to Eq. (\ref{ansatz_1}), 
we can approximate the ground state energy of the Hamiltonian by minimizing the following functional, which is expressed with the following inequality:
\begin{equation}\label{ritz}
	E_0 \le \frac{\langle{\Psi(\boldsymbol{\theta})}|\mathcal{H}|\Psi(\boldsymbol{\theta})\rangle}{\langle{\Psi(\boldsymbol{\theta})}|\Psi(\boldsymbol{\theta})\rangle}.
\end{equation}
The ansatz always gives an expected value that is greater than or equal to the ground state energy. 
However, if the chosen ansatz is a wave function orthogonal to the ground state wave function, then this method would give an estimate for one of the excited states of the system.

The general idea behind all of VQE-type algorithms boils down to the following. 
As a first step, it is necessary to define a target function  $C$, which encodes the solution to the problem. 
Then one have to choose a suitable ansatz depending on a discrete or continuous set of parameters $\boldsymbol{\theta}$. 
After that, one needs to perform optimization of the target function using a classical computer:
\begin{equation}\label{optimization_task}
	\boldsymbol{\theta}^{*} = \argmin_{\boldsymbol{\theta}}{C(\boldsymbol{\theta})}
\end{equation}

Considering the molecular ground state estimation problem, it is necessary to map the molecular Hamiltonian to the qubit Hamiltonian.
Under the Born-Oppenheimer approximation, the molecular Hamiltonian is usually expressed in its second-quantized form: 
\begin{equation}
	H = \sum_{p,q=1}^{M}h_{pq}a_{p}^{\dagger}a_{q} + \frac{1}{2}\sum_{p,q,r,s=1}^{M}g_{pqrs}a^{\dagger}_{p}a^{\dagger}_{r}a_{s}a_{q},
\end{equation}
where $a^{\dagger}_{p}(a_{p})$ is the fermionic creation (annihilation) operator and $M$ is the total number of SOs. 
The coefficients $h_{pq}$ and $g_{pqrs}$ are called one- and two-electron integrals and can be computed classically. 

To transform fermionic operators into qubit operators, several mappings, such as  Jordan-Wigner, Parity, and Bravyi-Kitaev, can be used. 
After the transformation is realized, one obtain a qubit Hamiltonian in the form: 
\begin{equation}\label{qubit_ham}
	H = \sum_{i} \beta_i P_i,  \quad \text{with} \quad P_i = \bigotimes_{k=1}^N \sigma_k^{(i)},
\end{equation}
where each $\sigma_k^{(i)} \in \{I, X, Y, Z\}$ is a Pauli operator, $N$ is a number of qubits, and $P_i$ are Pauli strings.
Each Pauli string can be then measured individually or in commutative groups using additional post-rotation gates, preceded by the same ansatz circuit for every Pauli string. 

The parametrized ansatz can be chosen based on a completely heuristic approach or on some physical intuition. 
In the case of quantum chemistry, we can adopt the well-known Unitary Coupled Cluster (UCC) ansatz, 
which can recover a portion of the electron correlation energy by evolving the Hartree-Fock initial wave function. 
Using the Hartree-Fock state as a reference, we can write the following expression:
\begin{equation}\label{exponential}
    |\Psi(\boldsymbol{\theta})\rangle = \hat{U}(\boldsymbol{\theta})|\psi_{HF}\rangle = e^{\hat{T}(\boldsymbol{\theta})-\hat{T}^{\dagger}(\boldsymbol{\theta})}|\psi_{HF}\rangle,
\end{equation}
where $\hat{T}(\boldsymbol{\theta})$ is a parametrized cluster operator and $\hat{U}(\boldsymbol{\theta})$ represents the corresponding unitary ansatz. 
Usually, only single-electron and double-electron excitations are considered, so: 
\begin{equation}\label{cluster_op}
	\hat{T}(\boldsymbol{\theta})=\hat{T}_1(\boldsymbol{\theta}) + \hat{T}_2(\boldsymbol{\theta})=\sum_{i} \theta_i P_i.
\end{equation} 

The algorithm for the classical optimization part of VQE should be chosen based on the specifics of a convergence, resource requirements, and noise tolerance for a certain problem. 
For instance, gradient-based optimizers like BFGS or L-BFGS are often preferred for smooth parameter landscapes due to their fast convergence~\cite{kandala2017}, 
while derivative-free methods such as COBYLA~\cite{cobyla} or SPSA~\cite{spsa_quantum} may better handle noisy quantum hardware at the cost of increased circuit evaluations. 
Additionally, the choice must balance computational overhead with scale of the problem, as iterative algorithms, such as ADAM~\cite{adam_vqe}, 
can adaptively adjust learning rates but may struggle with high-dimensional parameter spaces typical in large molecular systems.  

\section{VQE simulations}\label{sec:Resource}

Considering the aforementioned fermionic transformations and the STO-3G basis set, each SO can be encoded with one qubit.
Thus, the total number of qubits is equal to the number of SOs: $N_{q} = M$.
For formic acid and methylamine we have 17 and 15 MOs, which correspond to 34 and 30 SOs, or equivalently, to the number of qubits required for each system, respectively. 

The number of qubits can be reduced by employing the frozen core approximation, where the inner 1s orbitals of carbon and oxygen in the formic acid molecule and carbon and nitrogen in the methylamine molecule are frozen. 
This step reduces the required number of qubits by 6 for CH$_2$O$_2$ and by 4 for CH$_{3}$NH$_{2}$. 

Additionally, by applying a technique known as \textit{qubit tapering}~\cite{bravyi2017},
one can exploit symmetries of the system to find a unitary transformation $U$, which acts on the Hamiltonian (\ref{qubit_ham}) such that:
\begin{equation}
    H'=UHU^{\dagger}=\sum_i \beta'_i P'_i,
\end{equation}
where $H'$ has the same eigenvalues as $H$.
After this transformation, there are operators acting trivially in each Pauli string $P'_i$ of the transformed Hamiltonian. 
In other words, the Pauli strings have either the identity operator $I$ or the same Pauli operator (e.g., $X$) at specific qubit positions across all $H'$ terms. 
These operators can be replaced by their eigenvalues, and the corresponding qubits can be effectively removed from the simulation. In general, if there are $m$ independent $\mathbb{Z}_2$ symmetries, the number of active qubits is reduced by $m$.
In the case of formic acid and  methylamine molecules, we can eliminate three additional qubits from our simulation for both molecules, resulting in final counts of 25 and 23 qubits, respectively.

\begin{table}[h!]
\centering
\begin{tabular}{c c c c c}
\hline
Molecule & Atom & X (A) & Y (A) & Z (A) \\ \hline
\multirow{7}{*}{Methylamine} 

& C	& 0.0519020	& 0.7064670	& 0.0000000 \\
& N	& 0.0519020	& -0.7615000	& 0.0000000 \\
& H	& -0.9428060	& 1.1684160	& 0.0000000 \\
& H	& 0.5906740	& 1.0624810	& 0.8789330 \\
& H	& 0.5906740	& 1.0624810	& -0.8789330 \\
& H	& -0.4566340	& -1.1008410	& -0.8075530 \\
& H	& -0.4566340	& -1.1008410	& 0.8075530 \\ \hline

\multirow{5}{*}{Formic Acid}  
& C	& 0.0000000	& 0.3858930	& 0.0000000 \\
& O	& -0.8988900	& -0.6261750	& 0.0000000 \\
& O	& 1.1799510	& 0.1951720	& 0.0000000 \\
& H	& -0.4628290	& 1.3844990	& 0.0000000 \\
& H	& -1.7856570	& -0.2518330	& 0.0000000 \\ \hline
\end{tabular}
\caption{Calculated atomic coordinates for methylamine (CH$_3$NH$_2$) and formic acid (CH$_2$O$_2$) in angstroms. The coordinates sourced from the CCCBDB database~\cite{cccbdb} and optimized at the CCSD(T) level of theory with the Def2TZVPP basis set.}
\label{table:combined_coords}
\end{table}

Utilizing the coordinates in Table~\ref{table:combined_coords} allows us to derive the qubit form of the molecular Hamiltonian~(\ref{qubit_ham}) and ascertain the number of terms it contains. 
For molecular Hamiltonians it scales with the number of electrons $N$ as $\mathcal{O}(N^4)$.
This plays a central role in defining the required number of quantum circuits to be measured at each iteration and represents one of the major bottlenecks in the VQE implementation for complex molecules.
To partially overcome this problem one can adopt more sophisticated grouping methods for Pauli terms, such as \textit{qubit-wise} and \textit{general} groupings. 
These methods leverage the commutative properties of Pauli terms to notably decrease the number of necessary measurements, scaling as $\sim \mathcal{O}(N^4/30)$ for large Hamiltonians. We can say that two Pauli strings \textit{qubit-wise} commute if, at each index, the corresponding two Pauli operators commute, for instance, $\{YY, YI, IY, II\}$. In the case of \textit{general} commutativity, two Pauli strings commute if and only if they do not commute on an even number of indices, for example, $\{ XX, YY, ZZ\}$~\cite{gokhale2020optimization}.
Such sets can be measured simultaneously with a single circuit, leveraging their commutativity.

In considering the UCC Singles and Doubles (UCCSD) variational ansatz and employing the first-order Suzuki-Trotter decomposition, 
we aim to quantify the electronic excitations, the depth of the parameterized quantum circuit, and the total number of gates within these circuits.
If we consider a system with $N$ electrons and define the number of single excitations as $N_S = N(M - N)$, and the number of double excitations as:
\begin{equation}
    N_D = \frac{N(N-1)}{2} \,\frac{(M-N)(M-N-1)}{2}.
\end{equation}
Then the total number of one-qubit and two-qubit gates in the ansatz circuit can be approximately estimated as follows:
\begin{equation}\label{sd_num}
\begin{aligned}
	N_{\text{1-qubit}} &= \alpha_S \, N_S + \alpha_D \, N_D, \\
	N_{\text{2-qubit}} &= \beta_S \, N_S + \beta_D \, N_D.
\end{aligned}
\end{equation}
where
$\alpha_S$ ($\beta_S$) is the average number of one-qubit (two-qubit) gates per single excitation and
$\alpha_D$ ($\beta_D$) is the average number of one-qubit (two-qubit) gates per double excitation.

The total circuit depth can be estimated in a similar way:
\begin{equation}\label{tot_d}
	d = \gamma_S \, N_S + \gamma_D \, N_D,
\end{equation}
where $\gamma_S, \; \gamma_D $ are the contributions to the circuit depth from single and double excitations, respectively. 
The coefficients from Eqs.~(\ref{sd_num}) and (\ref{tot_d}) depend on the actual definition of excitation sub-circuits in the ansatz, the fermionic transformation used, and the number of qubits. 
Standard implementations of exponentials of Pauli strings (\ref{exponential}-\ref{cluster_op}) in quantum circuits typically rely on sequences of single-qubit rotations and cascades of CNOT gates arranged in a ladder-like structure.
Recent papers have presented optimal ways of constructing such circuits with fewer CNOT gates~\cite{yordanov2020efficient, yordanov2021qubit, magoulas2023linear}. 
One approach is to use \textit{qubit excitations}, which always require a fixed number of CNOT gates. 
However, it remains unclear whether they can adequately account for the electronic correlation contribution, as they ignore the anticommutation of fermions by excluding Pauli-Z products from the operator exponent.

\begin{table*}[t!]
\centering
\begin{tabular}{lcccc|cccc}
\toprule
\textbf{Parameter}                      & \multicolumn{3}{c}{\textbf{CH$_3$NH$_2$}} & \multicolumn{5}{c}{\textbf{CH$_2$O$_2$}}  \\ \hline \cmidrule(lr){2-4} \cmidrule(lr){5-8}
Ansatz Type                             & UCCSD  & k-UpGSD   & \textbf{UCCSDQs}    & k-UpGSDQ  & UCCSD  & k-UpGSD    & \textbf{UCCSDQs}     & k-UpGSDQ    \\ \midrule
Circuit Depth                           & 514379 & 14758   & \textbf{5766}       & 473     & 563722 & 15026    & \textbf{7563}        & 514         \\
Qubits                                  & 23     & 23      & \textbf{26}         & 26      & 25     & 25       & \textbf{28}          & 28          \\
Total Gates                             & 631022 & 18526   & \textbf{11558}      & 2822    & 682524 & 19466    & \textbf{14610}       & 3294         \\
Two-qubit gates                         & 457800 & 12896   & \textbf{4130}       & 1014    & 510426 & 13520    & \textbf{5221}        & 1183          \\
Pauli Terms                             & 20908  & 20908   & \textbf{20908}      & 20908   & 30423  & 30423    & \textbf{30423}       & 30423       \\
Circuits                                & 4144   & 4144    & \textbf{4144}       & 4144    & 5103   & 5103     & \textbf{5103}        & 5103        \\
Double Excitations                      & 2394   & 78      & \textbf{314}        & 78      & 2745   & 91       & \textbf{397}         & 91             \\ 
Tapering                                & Yes    & Yes     & \textbf{No}         & No      & Yes    & Yes      & \textbf{No}          & No             \\\bottomrule
\end{tabular}
\caption{Estimated resources for the VQE computation of CH$_3$NH$_2$ and CH$_2$O$_2$ molecules. The number of qubits is calculated using the frozen core approximation and the qubit tapering procedure. The number of circuits is based on qubit-wise operator grouping. The parameters for the k-UpGSD and k-UpGSDQ ansatzes are estimated with $k=1$, To obtain estimates for higher values of $k$, the table values should be multiplied by the corresponding scaling factor. The best-performing method in terms of the trade-off between required resources and VQE convergence to the FCI energy is highlighted in bold.} 
\label{tab:comparison}
\end{table*}

Alternative methods for constructing the variational ansatz, such as \textit{adaptive ansatzes} (ADAPT-VQE)~\cite{grimsley2019, Sapova2022} or \textit{evolutionary algorithms}~\cite{chivilikhin2020}, can reduce the computational resources, 
for example by shortening the overall circuit depth (\ref{tot_d}) or reducing the total number of variational parameters. 
However, they can incur a higher cost in terms of the number of circuit measurements, because each step in the adaptive or evolutionary procedure often requires extra loops and operator gradient evaluations.

A naive implementation of the UCCSD ansatz leads to excessive resource requirements since the circuit depth (\ref{tot_d}) is proportional to the number of two-qubit gates which scales as $\mathcal{O}(N^4)$.
However, one can employ several techniques to reduce these requirements. 
One of them is k-UpCCGSD (for brevity k-UpGSD)~\cite{Lee2018GeneralizedUCC, burton2023exact} ansatz,where $k$ represents the number of ansatz repetitions, each with a new set of variational parameters, $p$ refers to paired excitations, and $G$ denotes the inclusion of generalized excitations.
The main difference from UCCSD is that it includes only two-body terms, which shift pairs of opposite-spin electrons from fully occupied to completely unoccupied spatial orbitals making sure there are no singly occupied states described. The ansatz circuit's depth is reduced because the number of paired generalized excitations is significantly smaller than that of standard UCCSD excitations.

Another way to optimize the UCCSD ansatz is to filter out the excitation terms by their irreducible representations~\cite{cao2022}. 
In other words, one should retain only those terms whose irreducible representations coincide with the irreducible representation of the initial state (Hartree-Fock determinant), in the case of doubly occupied molecular orbitals we can write:
\begin{equation}
    \forall \Gamma \left(e^{\hat{T} - \hat{T}^\dagger} |\psi_{HF}\rangle \right) = \Gamma (|\psi_{HF}\rangle) : e^{\hat{T} - \hat{T}^\dagger} = 1,
\end{equation}
here $\Gamma$ denotes the irreducible irreducible representation of the corresponding wavefunction, ensuring that only excitations preserving the symmetry of the reference state $|\psi_{HF}\rangle$ are valid in the UCCSD ansatz.

\section{Results}\label{sec:results}
We propose several optimization strategies that realize different compatible combinations of the aforementioned methods.
Qubit excitations are designated as $Q$, and irreducible representation filtering as $s$, which are combined with UCCSD and k-UpGSD ansatzes to form UCCSDQ, UCCSDQs and k-UpGSDQ. 
Additionaly we make resource estimations for UCCSD and k-UpGSD ansatzes in combination with qubit tapering and compare them with other optimization strategies. In all estimates, we use \textit{qubit-wise} grouping of Pauli strings and an in-house quantum chemistry library specifically developed for the efficient implementation of variational quantum algorithms.~\cite{makushin2024quantum}.

First, to make sure the chosen ansatz construction methods and optimization strategies work as expected, we test them on two relatively simple molecules, LiH and BeH$_2$, both with a small number of SO (STO-3G basis suggested for both molecules). 
For LiH, we consider freezing the inner 1s orbital, which results in 10 qubits in the non-tapered regime and 6 in the tapered regime. 
For BeH$_2$, we initially have 14 qubits; the frozen core method reduces this to 12 qubits in the non-tapered regime and 7 qubits with tapering. 
We then calculate their ground state energies using the VQE and compare the results for different ansatz types. 
For both molecules, we use a noiseless statevector simulator in the combination with the BFGS algorithm for classical optimization.

\begin{figure}[h]
    \centering
    \includegraphics[width=1\linewidth]{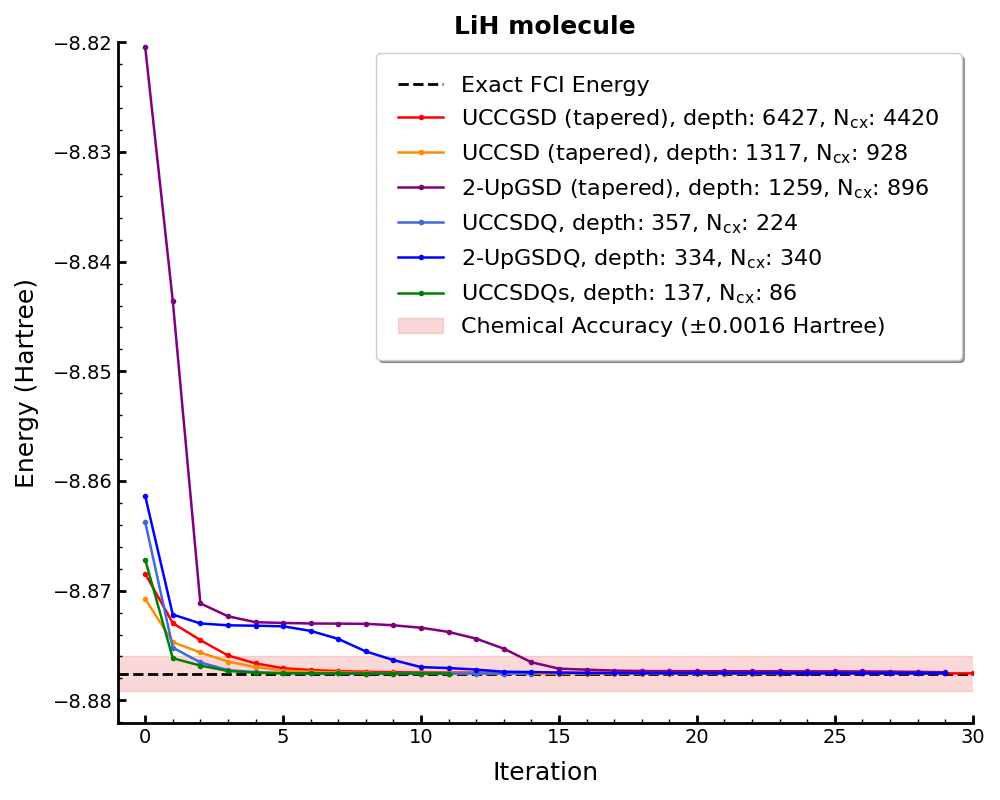}
    \caption{VQE calculation for the LiH molecule using various types of ansatzes and optimization methods. 
    The plot compares the convergence of energy values with the FCI energy as a reference and indicates the chemical accuracy threshold, N$_\text{cx}$ represents the number of two-qubit gates.}
    \label{fig:lih}
\end{figure}

The results of the VQE calculations for the LiH molecule are presented in Fig.~\ref{fig:lih}.
The plot shows how the energy converges over iterations for various ansatz types, such as UCCSD (plus its generalized version UCCGSD), 2-UpGSD, 2-UpGSDQ, UCCSDQ, and UCCSDQs. 
The dashed black line represents the exact FCI energy, which we use as a reference. 
The shaded area around it shows the chemical accuracy threshold of $\pm$0.0016 Hartree. 
Among all the ansatz types, UCCSDQs demonstrates the best performance: it achieves the lowest circuit depth (137) and two-qubit gate count (86), while still maintaining chemical accuracy.

\begin{figure}[h!]
	\centering
	\includegraphics[width=1\linewidth]{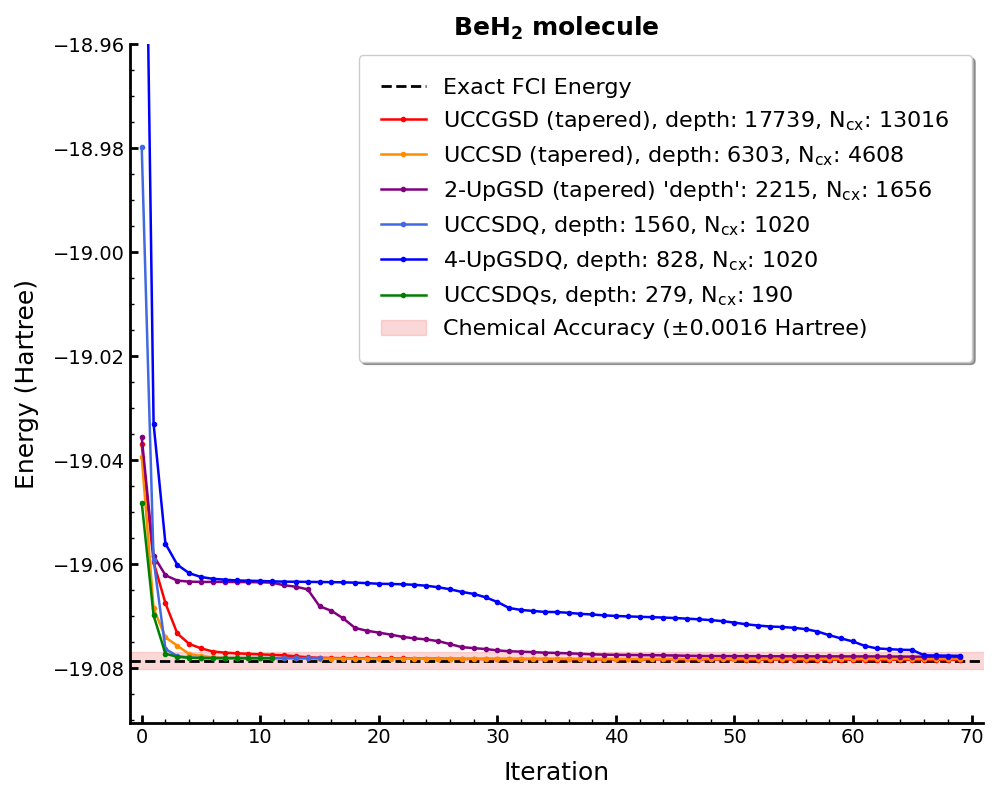}
	\caption{VQE calculation for the BeH$_2$ molecule using various types of ansatzes and optimization methods. 
	The plot compares the convergence of energy values with the FCI energy as a reference and indicates the chemical accuracy threshold, N$_\text{cx}$ represents the number of two-qubit gates.}
	\label{fig:beh2}
\end{figure}

In Fig.~\ref{fig:beh2}, we present the same analysis for the BeH$_2$ molecule. 
Again, one can see how each ansatz converges to the ground state energy. The k-UpGSDQ ansatz requires 4 circuit repetitions to achieve convergence.
As in the case of the LiH simulation, UCCSDQs proves to be the most resource-efficient, as it requires the least resources (circuit depth of 279 and two-qubit gate count of 190), while staying within the chemical accuracy range.
The energy curves initially drop sharply and then level off as they approach the exact FCI value, highlighting the efficiency of these methods for BeH$_2$.

Overall, these results show that the optimization strategies we use perform well for calculating ground state energies of small molecules like LiH and BeH$_2$. 
They also highlight how different ansatz types can affect the balance between accuracy and resource needs. 
While these strategies perform great for smaller systems, scaling them up to larger molecules is a whole different challenge.
Indeed, as shown in Table~\ref{tab:comparison}, which gives resource estimates for more complex molecules, such as methylamine and formic acid, the computational demands become significant. 
Even with the STO-3G minimal basis set, the required circuit depth and number of qubits for these molecules are way beyond what current quantum hardware can handle. 
\begin{figure}[h!]
    \centering
    \includegraphics[width=1\linewidth]{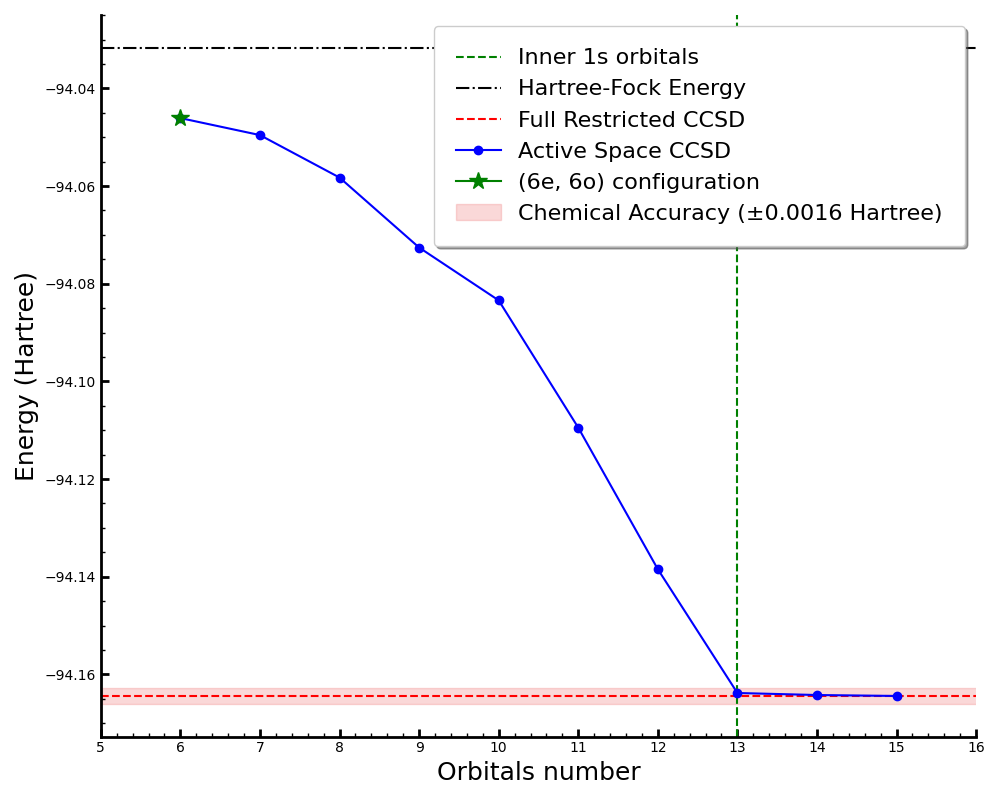}
    \caption{The ground-state energy for different active spaces of the methylamine molecule is shown. The green star indicates the active space configuration used in the VQE calculation. The green vertical line marks the active space region where only the core orbitals are frozen.}
    \label{fig:orbs_methyl}
\end{figure}
The use of more complex bases, that provide a better description of valence orbitals or account for polarization effects and electron correlations in molecules, such as 6-31G, 6-31G*, or cc-pVDZ, is associated with similar difficulties. 
All of them imply that significantly more qubits are required compared to the minimal basis case. 
For instance, simulating the H$_{4}$ molecule using the cc-pVDZ basis would require 40 qubits~\cite{xu2022}. 
For simulating the H$_2$O molecule in the 6-31g basis, 50 qubits are needed.
For methylamine and formic acid molecules the number of qubits exceed 60 in the case of complex bases.
Recent studies~\cite{Li2021, Shang2023} investigated various ways to enable the simulation of relatively large systems with small computational resources. A commonly used approach in these works is the HOMO-LUMO approximation, which restricts the active space to a subset of orbitals of a molecule. This method results in a more compact qubit representation of the molecular system, making simulations more feasible. However, this reduction in active space introduces additional approximation errors, which can impact accuracy.

\begin{figure}[h!]
    \centering
    \includegraphics[width=1\linewidth]{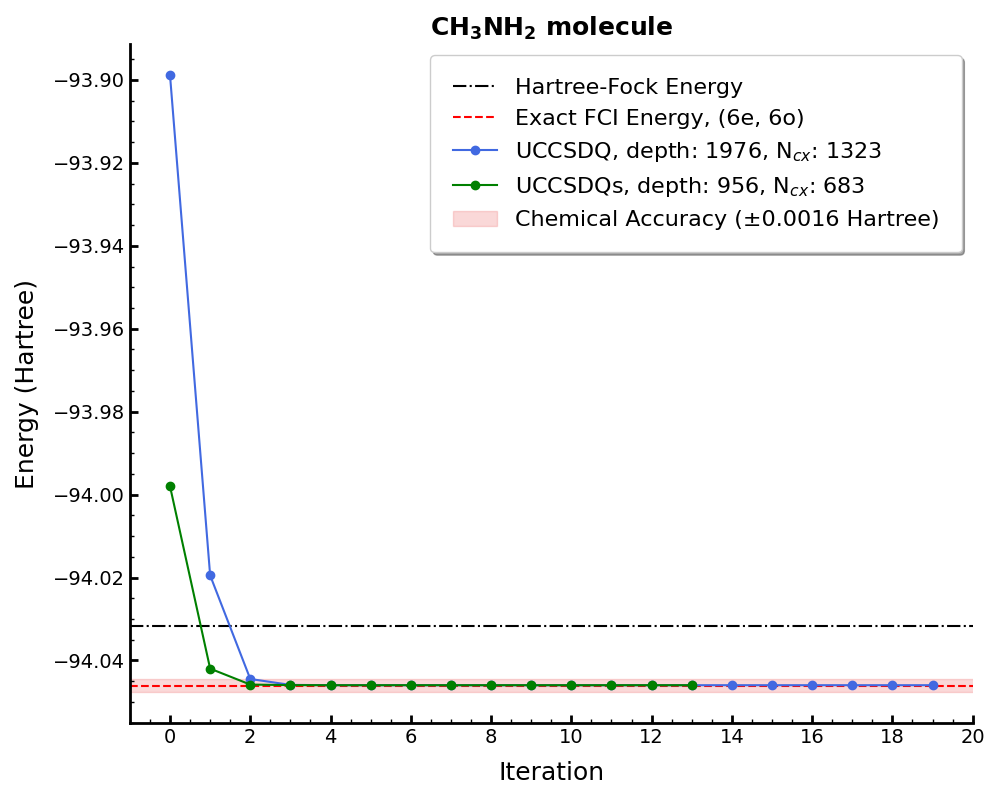}
    \caption{VQE calculation for the CH$_3$NH$_2$ molecule in (6e, 6o) active space using
UCCSDQ and UCCSQs ansatzes. The plot
compares the convergence of energy values with the FCI en
ergy as a reference and indicates the chemical accuracy threshold, N$_\text{cx}$ represents the number of two-qubit gates.}
    \label{fig:methyl_vqe}
\end{figure}

Similarly, we demonstrate the results of simulations for the Methylamine molecule with a restricted number of spin orbitals. 
We take an active space of 6 electrons and 6 spin orbitals from the full 15 spin-orbital space. 
First, we will show the energy difference for such a choice. In Fig.~\ref{fig:orbs_methyl}, one can observe different configurations of the methylamine active space and their corresponding energies. 
This choice of active space reduces the number of qubits required to simulate the molecular system.
However, as the number of spin orbitals decreases, the calculated energy becomes closer to the Hartree-Fock result. 
Chemical accuracy is achieved only when the core 1s orbitals are frozen.
For the (6e, 6o) active space configuration of methylamine we use the UCCSDQ and UCCSDQs ansatzes in VQE calculation and the same minimal basis STO-3G. 
As it shown in Fig~\ref{fig:methyl_vqe} the energy converges to the FCI value for both ansatzes. 
This result demonstrates that for the 12-qubit methylamine active space, the VQE algorithm, combined with the aforementioned absatzes, can achieve chemical accuracy within a small number of iterations.
Other types of ansatzes are not tested due to higher computational cost.

The VQE calculation for the formic acid molecule in a reduced active space, similar to the one performed for methylamine, does not show good convergence to the expected ground state energy. The reason might lie in the stronger electronic correlations between carbon and oxygen atoms in the formic acid molecule, which could not be adequately modeled by circuits based on \textit{qubit excitations}. It is possible that effective fermionic excitation circuits, as proposed in Ref.~\cite{yordanov2020efficient}, could achieve better performance, however at the cost of an increased two-qubit gate count.

\section{Conclusions}\label{sec:conclusion}

Quantum chemistry is one of the most promising avenues for applications of quantum computing devices, with the VQE and its various extensions currently serving as the primary methods for quantum chemical simulations.
In this work, we present the results of a computational resource analysis for organic molecules, such as formic acid and methylamine, within the VQE framework.
Additionally, we investigate several optimization strategies to achieve better quantum resource scaling.
To the best of our knowledge, no prior studies have systematically explored the combined effect of these optimization strategies on VQE performance.

 To demonstrate their adequacy, we calculate the energy spectra of small molecules, such as LiH and BeH$_2$, as well as methylamine with a restricted active space of spin orbitals, all in the minimal basis.
Our analysis show that the combination of \textit{qubit excitations} with filtration by irreducible representations (UCCSDQs) could drastically decrease the ansatz depth and the number of two-qubit gates, 
while still recovering the major part of the correlation energy.

We estimate the quantum resources required for full active-space configuration simulations of the methylamine and formic acid molecules using different types of ansatzes and present the results in Table~\ref{tab:comparison} for comparison.
We highlight the UCCSDQs results as most promising optimization strategy in terms of the circuit depth and algorithm convergence.
Combining UCCSDQs with the \textit{qubit tapering} procedure could lead to further improvements; 
however, it requires a more careful analysis of the connection between $\mathbb{Z}_2$ symmetries and the representation of the excitation operator circuits, which is beyond the scope of this study. 
The k-UpGSD ansatz may result in smaller improvements, but at the same time, it can be easily combined with \textit{qubit tapering}. 
A possible challenge for this ansatz is that, as the number of qubits increases, the process of classical optimization may become stuck in local minima. 
In such cases, this issue could potentially be mitigated by increasing $k$, i.e., by increasing the number of variational parameters. For example, in the BeH$_2$ calculation (see Fig.~\ref{fig:beh2}), we used $k = 4$ to achieve convergence.
This, however, comes at the cost of increased circuit depth, which amplifies the impact of two-qubit gate errors.

The quality of two-qubit operations in quantum computing devices is a crucial bottleneck for the real-world application of quantum chemical simulations.
As a result, ansatzes that require a large number of such operations become impractical, necessitating the search for optimization strategies.
We have considered possible ways for  overcoming these challenges by using various combinations of previously developed optimization methods. 
However, challenges remain, particularly for larger molecules and more complex systems.
We expect that further improvements in ansatz design, gate fidelity, and error mitigation may greatly lessen these restrictions and open the door for practical applications of quantum algorithms such as VQE in quantum chemistry.

\textbf{Data availability}.
The data supporting the findings of this study are available from the corresponding
authors upon reasonable request.

\textbf{Acknowledgments}. 
A.K.F. acknowledges the support from the Priority 2030 program at the NIST “MISIS” under the project K1-2022-027.

\textbf{Competing interests}.
Owing to the employments and consulting activities of K.M.M. and A.K.F., they have financial interests in the commercial applications of quantum computing.

\bibliography{bibliography}
\end{document}